# EURYDICE

## A platform for unified access to documents

Serge Rouveyrol, Yves Chiaramella, Francesca Leinardi, Joanna Janik
(IMAG – Institut d'Informatique et Mathématiques Appliquées de Grenoble http://www.imag.fr)

Bruno Marmol, Carole Silvy, Catherine Allauzun
(INRIA Rhône-Alpes http://www.inrialpes.fr)

14 october, 2001

## ABSTRACT

In this paper we present Eurydice, a platform dedicated to provide a unified gateway to documents. Its basic functionalities about collecting documents have been designed based on a long experience about the management of scientific documentation among large and demanding academic communities such as IMAG and INRIA. Besides the basic problem of accessing documents – which was of course the original and main motivation of the project - a great effort has been dedicated to the development of management functionalities which could help institutions to control, analyse the current situation about the use of the documentation, and finally to set a better ground for a documentation policy. Finally a great emphasis – and corresponding technical investment – has been put on the protection of property and reproduction rights both from the users' intitution side and from the editors' side.

Developers staff :
    Serge Rouveyrol, Cyril Araud (IMAG)
    Bruno Marmol, Florian Dufour (INRIA Rhône-Alpes)

# I. Introduction

Electronic access to information has been a main concern since the nineties. It has led to a considerable improvement in document exchanges between libraries and to an equally important evolution considering users' autonomy. Electronic access is by now a common feature of all research and university libraries and has triggered a fast evolution of these institutions.
This authentic mutation has occurred in the context of fast development and dissemination of new technologies but also often in the context of a relative stagnation - when not a decrease - of financial resources. The access to scientific information has by now a considerable impact on library budgets. Considering either simple operations such as information retrieval (access to databses) or access to documents themselves (either in paper or in electronic formats), subsription costs have more than doubled (particularly in the domain of exact sciences). It is by now a well-known concern that the dependancy about editors may increase with the generalisation of electronic documents compared to the current tradition of paper documents.

Warned by information professionals, universities and research institutes of the Rhône-Alpes region have decided to promote a long-term policy for a better exploitation of the regional documentary resources. This policy is clearly inspired from examples in the US and Canada.

### 1.1. BRAIN and CAMPRA

The BRAIN[1] initiative was launched in 1998. To undertake this project, CURA[2] - a consortium of regional universities and french Grandes Ecoles, along with national research institutions such as CNRS[3], Cemagref[4], INSERM[5], INRIA[6], INRETS[7] has decided to found its action on a coordination among all these involved institutions. Five working groups have been designed about (i) access to electronic journals and catalogs (ii) acquisition of digitalised corpuses (iii) digitalisation of paper documents (iv) digitalisation of manual catalogs and (v) the design of a regional gateway on the web for accessing electronic documentation. All the proposed actions were aimed to benefit to the entire community of the researchers of the region.
The paraticipant institutions have – among other specifications – aimed to the optimization of subscription funds use the amount of which cumulates to several tens on million francs at the regional level. This optimization is undertaken via a common organisation regarding negociations with editors to obtain better prices, and with the Centre Français de la Photocopie regarding duplication rights. The creation of a regional consortium quickly appeared as an obvious solution as the infrastructure of an effective, multi-site, coordination.
According to this policy, CAMPRA[8], a regional consortium of documentation centers supported by BRAIN has been designed and assigned three majors goals :
- distributed access to large electronic catalogs such as Current Contents, Inspec etc.
- access to electronic versions of journals subscribed by the consortium.
- design of a platform aimed to support a common policy about the mutualisation of electronic documents and a better acquisition strategy.

The design, implementation and experimentation of the Eurydice[9] platform is the main outcome of the third axis of this regional policy.

### 1. 2. Outlines of the Eurydice Project

Eurydice is a platform aimed to provide all the partners of CAMPRA a unique environment for accessing documents (either in paper or electronic format) and electronic catalogs.
This ongoing project is financially supported by the Rhône-Alpes region and the participants involved in the project on a 50% basis. A first phasis of the project has led to the development of a first version of the platform and involved three partners : IMAG[10], INRIA Rhône-Alpes and XRCE[11]. This original version was derived from a former platform dedicated to the digitalization « on demand » and the electronic distribution of paper documents named CALLIOPE[12]. Two new partners – ISH[13] and ENSL[14] - have recently joined this development group for an experimentation program which constitutes the second phasis of the project. This second phasis has recently started with the installation of the Eurydice platform in these two institutions, and its outcomes – through extensive final-users tests – will hopefully ease the dissemination of the platform among the regional members of the CAMPRA consortium.

---

[1] BRAIN :Bibliothèque Rhônalpine de l'Information Numérique
[2] CURA : Conférence Universitaire Rhône-Alpes
[3] CNRS : Centre National de la Recherche Scientifique. URL : http://www.cnrs.fr
[4] Cemagref : La Recherche pour l'Ingénierie de l'Agriculture et de l'Environnement. URL : http://www.cemagref.fr
[5] INSERM : Institut National de la Santé et de la Recherche Médicale. URL : http://www.inserm.fr
[6] INRIA : Institut National de la Recherche en Informatique et en Automatique. URL : http://www.inria.fr
[7] INRETS : Institut National de Recherche sur les Transports et leur Sécurité. URL : http://www.inrets.fr
[8] CAMPRA : Consortium pour l'Accès Mutualisé aux Publications scientifiques numérisées en Rhône-Alpes
[9] Eurydice : épouse d'Orphée, bru de Calliope
[10] IMAG : Institut d'Informatique et Mathematiques Appliquées de Grenoble. URL : http://www.imag.fr

## II. The Eurydice Platform

*2.1. Main functionnalities*

Eurydice is a platform managing controlled access to scientific publications. The aimed users are researchers and academics from the member institutions participating in the experiments and – ultimately - any institution within CAMPRA. The general goal is to provide a distributed access to papers within journal collections. An important feature lies in the possibility for any registered user to access to papers from journals which have been subscribed by his institution, or by any other associated institution.
Eurydice has been designed to provide access to articles themselves and to a series of services integrated in a unique interface, based on pre-defined institution profiles. Access rights and associated services are related to these profiles.

*From users' point of view*

The starting point when using Eurydice is the Summary Server which contains lists of journals subscribed by each institution. Journal titles are clustered based on domains. By now – during the experiment phasis – two domains have been selected : human sciences and exact sciences (including computer science and applied mathematics), any given title being possibly related to one or two domains. The selection of relevant titles to be offered in this context results from a concertation between the various documentation centers associated within the consortium, depending on their actual subscriptions and the services they plan to provide (see below the corresponding constraints).

Access to Eurydice is controlled, based on user identification and/or user's machine IP number. This allows for adapting formats of displayed pages and services to the actual rights subscribed by each user's isntitution, regarding accessed information sources. As will be seen later, this distinction is shown to the users based on specific access icons displayed jointly with document titles.

Beyond the simple browsing through titles lists sorted by domains or by journal titles, Eurydice provides – and this is a main feauture of the platform - a direct access to any individually selected paper.
Downloading selected papers is not systematic ; this is done only at the explicit request of the authorized user (ie. if the user estimates that the article is relevant to his/her information need). The downloading process itself is optimized by a cacheing strategy which avoids further network accesses to the paper home-site if an other authorized user wants to download it (once a paper has been dowloaded, a specific icon signals that it is now fast-accessible on the site for further authorized accesses). When selecting this second case by depressing the corresponding button, the user will trigger – in a totally transparent way – a procedure adapted to his own access rights (ie. the access rights subscribed by his mother institution) and to the availability of the requested article.

The available procedures are described in the following table :

---

[11] XRCE : Xerox Research Center Europe. URL : http://www.xerox.fr/
[12] Cf . : Alauzun, C. et Laurent, P. - « CALLIOPE : un exemple de mise à disposition d'un fonds documentaire scientifique », *Micro-bulletin*, 1999, 3.
[13] ISH : Institut des Sciences de l'Homme de Lyon. URL : http://www.ish-lyon.cnrs.fr/
[14] ENSL : Ecole Normale Supérieure de Lyon. URL : http://www.ens-lyon.fr/

| *Original fomat* | Origin | Processing | | | Paper Destination | Paper Delivery | *Reception format* | |
|---|---|---|---|---|---|---|---|---|
| *Electronic* | The institution subscribes to the electronic version of the journal | Storage of the original pdf format on the document server of the subscribing institution | Transfert of the original pdf format | | Member of a subscribing institution | Delivery on the user's workstation | *Electronic* | **Local Mode Access** |
| | | | Printnig by the subscribing institution | | Member of an affiliated, but not subscribing institution | Delivery on an authorized printer of the affiliated institution | *paper* | **Shared Mode Access** |
| *paper* | The institution subscribes to the paper version of the journal | Digitalisation of the article from its paper format | Storage on the document server of the subscribing institution | Printing the document server of the subscribing institution | All authorized users | Delivery on an authorized printer | *paper* | |
| *paper* | The institution subscribes to the paper version of the journal and cannot digitalize paper-version articles | Classical photocopy | | | All authorized users | Delivery by postal mail | *paper* | |

Beyond this basic functionality about accessing documents, whatever their original format through a single, unified procedure and a unique environment, Eurydice offers :
- a search engine to investigate the Summary Database
- an Alert Service about one or several titles selected by each individual user

It should be clear however that Eurydice is not a bibliographic database. Information available on the Summary Server is extracted from journal summaries : title and journal ISSN, title and authors of articles, abstracts (when available), volume number, date and page numbers. These are the information items indexed by the system and hence usable as query items for the search engine (based on a simple query interface).

Summaries are stored on a common server deserving for all the affiliated partners. This centralised organisation allows Eurydice to provide the Alert Service mentioned above. This Alert Service allows any individual user to « subscribe » to any title, or list of titles, whatever the involved editors, and to receive (at his e-mail address) the last available summary(ies) as soon as they have been received from the editors (the summary database is updated on a daily basis). The user may then, from his own workstation, seek and access relevant papers immediatly after their electronic publication. By now the Alert Service is based only on title selection ; in the next future it is planned to base this filtering process on *thematic profiles* defined by individual users.

### *Eurydice and authors' copyrights*

Eurydice's underlying strategy about legal rights is based on french regulation in the domain of author copyrights and photocopy regulation. It should be clear now (from the table above) that whatever the situation, any requested article is delivered to the user using only two possible formats :
- if the user's mother institution has subscribed for the electronic version of the corresponding journal, the article is delivered in its electronic version on the user's workstation. The user may then store it or print it ; all legal aspects regarding this situation have been fixed when the mother institution has negociated the subscription with the editor, and hence are applicable to any member of this institution.
- If the mother institution has not subscribed either the paper or electronic version of the journal, the article may be delivered (if available) from an other affiliated institution, provided that it has subscribed to this journal.

  It is important to notice that in this case the *article is always delivered in paper format*. As shown in table 1 above, this may be done by printing the paper on a printer located in the user's institution, or by classical photocopy from a paper version of the journal and postal mailing to the user. The new situation here is that all these underlying procedures of requesting a printed version from an other affiliated site having the needed paper in electronic version, or asking an affiliated site having a paper version (when no electronic version is available among the affiliated institutions) are completely automatic and transparent for the user. The shape of the request icon warns him that his request can or cannot be satisfied immediatly and, in the later case, that he will receive a message advicing him when the article is available and through which procedure (ie. printed by an authorized document server on an authorised printer within his environment, or photocopied and mailed). When printed by a document server, the server accounts the copy and a trace of the procedure is stored for further internal billing procedure between the involved institutions .

  Most important regarding copyright is the fact that a payment record is automatically generated to fulfill legal copyright obligation to be paid to the CFC[15]. Finally, it is important also to notice that the user does not have to worry about which institution to contact to obtain the article : Eurydice entirely manages this aspect of the « localisation » process.

### *Libraries' Point of View*

As explained before, Eurydice provides direct access to document articles requested by individual users from their office. One might then see Eurydice as a powerful shortcut of library classical services and information circuits. It has to be pointed that using Eurydice, each library - or documantation center - keeps full control about its documentation policy, and particularly about the modalities (among those listed before) which will be offered to the users. Except for specific agreements between institutions willing to share information and services (and possibly costs), every journal subscription paid by a given documentation center of course remains its own property and in its complete access privacy.

---

[15] CFC : Centre Français d'exploitation du droit de Copie. A legal, nationwide institution, which collects copy taxes and redispatches due fees related to photocopies of private material to the editors. See the CFC web site at :
*http://www.cfcopies.com/accueil.htm*

On the other side, the documentation center may decide to *share information* (either in paper or in electronic format) and services with some other partner institution(s). It should be clear from the previous explanations that such a process will be (i) entirely based on paper versions of any requested article and (ii) that most of the administrative burden related to this kind of exchange will be undertaken by Eurydice itself. What remains to manual processing is photocopying or digitalizing the article (and mail it, in the later case).

Of course choosing the digitalization (upon mailed request) of documents as an offered service has potentially a strong impact on the organization and the work load of any voluntary library or documentation center. This service can be appropriately billed and managed using Eurydice, and collected fees may help to support the corresponding additional investment and salary costs. One main goal of the ongoing experiment is precisely to analyse - in real situations - the organisational and cost impacts of such approaches.

Besides friendly interfaces for collecting users' requests also available of course to librarians and their technical staff, Eurydice offers powerful management tools for maintaining document corpuses and for analysing users' behaviour and needs about documentation.

Each library or documentation center has to undertake a basic processing on available document summaries.

Though new summaries are automatically included in the summary database from various private – or possibly public – sources (eg. Swetscan, editor servers etc.), and despite the professionalism of these providers, errors or missing information within delivered summaries are not rare events. As a result, and in order to maintain information consistency, each partner has to control and maintain in a correct and up-to-date status the information provided on the Summary Server.

To fulfill all these important aspects of documentation management Eurydice provides an « Administration Interface» which allows documentation administrators to correct summaries, to manually add some missing ones, to modify access rights, to interrupt or suppress inactive or useless Alert Services etc.

We have already suggested the possibility of analysing users' behaviour through the unified interface of Eurydice. It is clear that Summary Servers and Document Servers keep records about accesses to their own categories of information (summaries, actual papers and journals). This information – of course kept anonymous in terms of individual identity – establishes an extremely valuable statistical link between institutions (ie. universities, schools, research labs, reseach groups, etc. all identified through lists of IP addresses) on one hand, and titles and journals (and hence on editors) on the other hand. Eurydice allows to maintain such statistics, and offers a fairly simple interface for exporting this raw data to an Excel™ application for example. Then it is up to the documentation manager to analyse, interpret and exploit this information for a better control of his/her documentation policy.

Finally it has to be noted that Eurydice allows the storage of all the information managed by the Summary Server, independantly of any commercial provider. This is an important feature : Eurydice then ensures the persistence of acquired summaries based on archives, and hence ensures their availbility to the users throughout time.

### *Institutions' Point of View*

Using Eurydice, libraries and documentation centers may collect accurate statistical information about information needs and users' behaviour, and turn them to their mother institutions. This information will in turn help institutions to be accurately informed about this important side of academic life and hence it will help making appropriate decisions in this domain. Moreover, these statistics will considerably help institutions to find and design documentation partnerships with other institutions owning complementary information sources, based on such accurate view of their users' needs compared to their available documentation and financial resources. This aspect has still to be experimentally evaluated, but there is some clear evidence that Eurydice could help reducing costs based on a better mutualisation of documentation costs.

## III. The System Structure

For historical reasons related to Calliope (see the introduction) the programming language is still Perl. The platform then is mainly based on free softwares (FreeWais, Apache, module CPAN[16]) and on

---

16 CPAN : Comprehensive Perl Archive Network

Webdoc™, a software developed by Xerox and dedicated to the management of document databases. Underlying systems are Solaris ans Linux and the adaptation of Eurydice under other systems should be simple as long as the underlying softwares remain available.

There is a kind of dual relationship between the software and its supporting hardware. A server is needed for hosting the Summary Server, and there may be only one Summary Server at a time. On the opposite there may be as many Document Servers as needed, all running under Webdoc. The standard configuration turns out to reduce to one Document Server by affiliated institution and only one Summary Server for a whole consortium.

Eurydice is made up of several components :

- the first – and most visible to the user – is the Summary Server. This server manages several of the most important functionalities provided by Eurydice, namely the summary database, the search engine for retrieving relevant titles, all the access controls, the Alert System for example.

- The second component is made of all the Webdoc Document Servers developed by Xerox Company. These servers manage the storage of documents that have been digitalized within affiliated document centers or collected from an editor's site. Every documentation center subscribing to electronic journals or willing to digitalize paper journals needs a Document Server. In this way, every electronic document physically remains in its own institutional – and legal considering property – environment.

- The third component is named « the Binder ». Given a user-selected reference from he Summary Database, this component receives via the HTTP protocol a request containing all the meta-information describing this document (title, editor etc.) and – from this information – dynamically searches the url of the corresponding electronic paper on its editor site. Hence the name « Binder » because it binds a meta-definition of an article (ie. its bibliographical reference) to the actual full-text electronic version of the referred article. Once the url found, the Binder communicates this information to the involved Summary Server. A direct access to the paper and its transfer to a Document Server is now feasible. One have to notice that this strategy implies the design of a Binder for each editor because they each have their own policy about document meta-data and electronic access.

In the remaining of this paper we shall now focus on the Summary Server and on the main Access Control Engine which is warrants proper protection of author and editor rights.

## IV. The Summary Server

The Summary Server is made of several components :

### 4.1. Modules dedicated to downloading and processing new summaries

Summaries may be collected from various sources : databases such as Swets/Blackwell, Ebsco or INIST[17], or directly from the Alert Service of the editors. The summary formats being different for each editor, a specific module has to be designed for each editor. These modules transform the original summaries into a pivot format (in XML) which is then the only format known by the Summary Server. The XML format was a natural choice for two main reasons : (i) immediate integration of international constraints and standards (use of Unicod for language-specific characters) and (ii) the vailability of numerous programming tools for this language. One of course could add that XML being an emerging standard, this choice could help exchanging references between institutions or reusing owned references on new implementations of the Summary Server.

### 4.2. Service Modules

These components process incoming summaries. A service module needs to be « registered » by the Service Manager to be activated. Once a Service Module is registered, the Service Manager may activate it through a specific API[18]. In this way new Service Modules may be added and executed without any modification of the code.

---

[17] INIST : Institut National d'Information Scientifique et Techn ique. A nationwide documentation institution managed by CNRS.
[18] API : Application Program Interface

- **a. Control Modules**

    A Control Module executes consistency controls such as checking either a given title is coherent with the given editor and the given ISSN. It cannot of course control titles and corresponding authors. This is an example of what was said above about the role of documentalists for maintaining consistency within summaries. Remember also that Eurydice provides an interface (part of the Administration Interface) to make these corrections.

- **b. Database Storage Module**

    This module loads the Summary Database by storing the summary and its associated meta-data in a file hierarchy. Not all transferred summaries are stored within the Summary Database ; a selection is done based on a configuration (a kind of filter) associated to each summary provider, or based on specific choices done by documentalists or users. From the system's point of view however, there is no theoretical limit to the number of summaries in the database ; the actual limits are set by the users an their domains of interest. Such domains being subject to evolution, the system allows to add new titles at any moment to editor configurations because the systems always keeps track of every information provided, or which have been provided in the past, by all editors. This information, stored in the original format of each editor, may be reused and processed at any moment, using the Control an Database Storage modules.

- **c. Alert Modules**

    The Alert Module uses the electronic mail to send to each user the last summaries corresponding to their « subscription » (see above). It is important to notice that the list involved in any user's « subscription » may contain titles being (at the moment) out of the actual subscription lists of the Document Server (ie. the full-text document is not accessible on this Document Server). This is possible because, as said before, the Summary Server stores everything coming from the summary editors (hence also summaries of journals not subscribed at the moment by the institution) ; This is certainly also useful in that access statistics on the Summary Server may in this way new potential information needs – or evolutions of domains of interest – which are not currently covered by the institution.

The following figure illustrates the relationship among all these components :

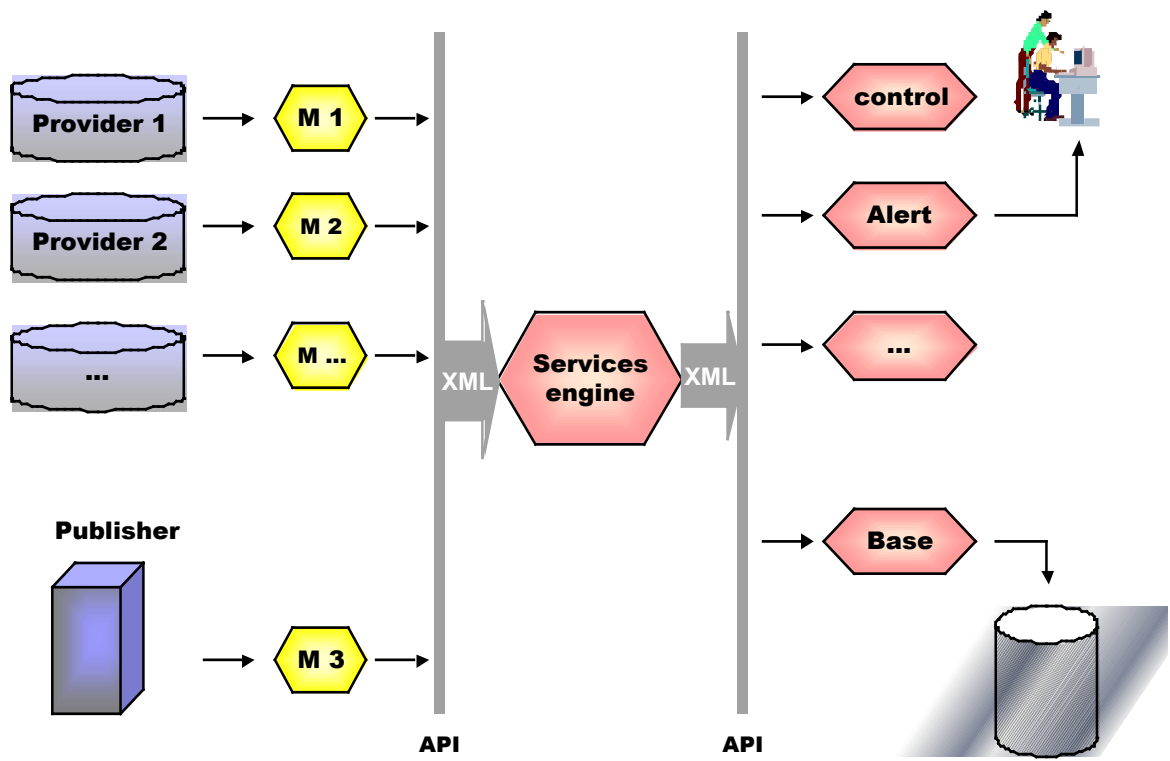

### *4.3 Interfaces*

Interface design is an important feature of such interactive environments and a great effort has been dedicated to this aspect.

#### *a. User Interfaces*

The user interface conforms with the recommendations of the W3C[19], and to the notion of accessibility as defined by the WAI[20].

#### *b. Administrators*

As said before, administration capabilities have been designed and implemented as a needed complement to the access interface. It includes the correction of summaries, data management, access to statistics and the modifications of configurations. Access to these operations is protected by the protection mechanism of the Web server itself. By now there are only two protection levels : (i) the users who are not allowed to modify configurations and (ii) the administrators (ususally from the library staff) who are allowed to modify any configuration.

#### *c. Search Modules*

Searching with Eurydice is performed based on FreeWais-sf, a free softwre for indexing and on Sfgate for managing the Web interface. Searching is performed only on titles and/or author names known in the summary database and not on the whole set of loaded summaries (remember that some of them may have not been loaded in the database).

---

[19] W3C : World Wide Web Consortium. URL : http://www.w3.org/
[20] WAI : World Accessibility Initiative. URL : http://www.w3.org/WAI/

### d. Communications with other servers

Communications with external software components (ie. Binder, Document Server) are based on the HTTP protocol and the XML document format. The information flow between the various elements is illustrated by the figure given below.

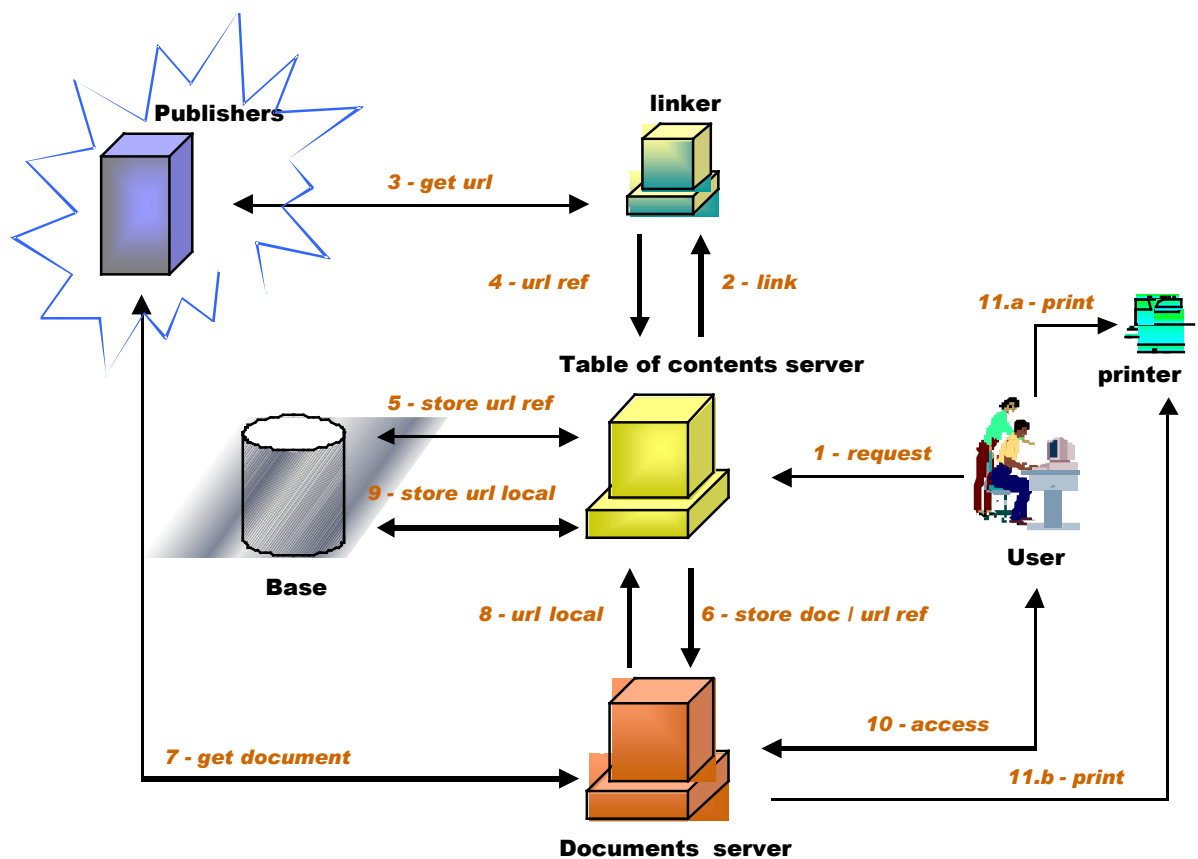

### e. Control of Access Rights

Effective control of access rights is a complex and fundamental task of the Eurydice platform. The task is fundamental – one might say « critical » - due to the close relationship between access and reproduction control and legal aspects implied by subscription contracts issued with the editors. Eurydice clearly aims at monitoring these two basic functions related to the use of documents by a community of users. Then its reliability in terms of property rights protection may be viewed as critical by any institution, and we may say that Eurydice has been precisely designed and implemented with this aspect as a primary concern. Security is by now based of a tight management of IP address lists and their relation with contractual subscriptions with editors regarding access and reproduction rights (again this is defined in the context of the french legislation in the domain). As explained before, accurate records are maintained about accesses, transfer and reproduction operations which in turn allow (in our opinion) *a priori* and *a posteriori* control about protection of property rights. In the fast evolving context which characterises the present situation of electronic publishing we think that this approach constitutes an acceptable compromize for most institutions and editors.

Technically speaking, access control is a difficult notion due to the number of parameters involved : IP numbers, configurations, or adaptation of access rights to specific user classes). At the moment, individual control based on login authentification seems irealistic due to the complexity and dynamicity of the covered populations of users (clear and relatively stable for academic staff, much difficult to monitor in real time for student populations).

In the present configuration of Eurydice, each documentation center defines its own policy about the services offered through Eurydice and manages its own list of IP addresses. This includes the definition of user categories and corresponding services considering the whole collection (ie. the set of all the subsribed journals).

The list of available services covered by these control procedures is by now :

- navigation and browsing
- alert service
- photocopy service
- digitalization
- electronic access
-

## V. Conclusion

Eurydice is a platform dedicated to provide a unified gateway to documents. Its basic functionalities about collecting documents have been designed based on a long experience about the management of scientific documentation among large and demanding academic communities such as IMAG and INRIA. Besides the basic problem of accessing documents – which was of course the original and main motivation of the project a great effort has been dedicated to the development of management functionalities which could help institutions to control, analyse the current situation about the use of the documentation, and finally to set a better ground for a documentation policy. Finally a great emphasis – and corresponding technical investment – has been put on the protection of property and reproduction rights both from the users' intitution side and from the editors' side.